\documentstyle[prb,multicol,aps,epsfig]{revtex}

\begin{document}
\title{Single parameter scaling in 1-$D$ localized absorbing systems}
\author{Lev I. Deych$\dagger$, Alexey Yamilov$\ddagger$, and A. A. Lisyansky$\ddagger$}
\address{$\dagger$Physics Department, Seton Hall University, South Orange, New Jersey 07052\\
$\ddagger$Physics Department, Queens College of CUNY, Flushing, New York 11367}
\date{\today}
\maketitle

\begin{abstract}
Numerical study of the scaling of transmission fluctuations in the 1-$D$ localization problem in the presence of absorption is carried out. Violations of single parameter scaling for lossy systems are found and explained on the basis of a new criterion for different types of scaling behavior derived by Deych et al [Phys. Rev. Lett., {\bf 84}, 2678 (2000)].
\end{abstract}

%\pacs{23.23.+x, 56.65.Dy}
%\begin{multicols}{2}

\section{Introduction}

The single parameter scaling (SPS) hypothesis is the cornerstone of the current understanding of the localization phenomena. It was originally formulated in terms of the scaling behavior of the conductance of disordered conductors,\cite{GangOfFour} where it was suggested that when the length, $L$, of a disordered conductor increases, the evolution of the conductance, $g(L)$, is determined by a single parameter, $g$ itself. For one-dimensional systems Landauer's formula \cite{Landauer} expresses the electron conductance in terms of electron reflection, $R(L)=rr^{*}$, and transmission, $T(L)=tt^{*}$, coefficients as $g(L)=R(L)/T(L)$, and thereby allows the considering of electron transport on an equal footing with these propagation of, for example, light. It was recognized later that SPS must be understood in terms of properties of the entire distribution of $g$ (or $T$), and that the most appropriate quantity to deal with is \cite{Anderson}
$$
\tilde{\gamma}(L)=(1/2L)\ln(1+1/g(L))=(1/2L)\ln(1/T(L)).
$$
In the limit of large $L$, this parameter  is normally distributed with the average $\gamma=\langle\tilde{\gamma}(L)\rangle=\lim_{L\rightarrow \infty}\tilde{\gamma}(L)$ and the variance $\sigma^2(L)=\langle\tilde{\gamma}^2(L)\rangle-\gamma^2$. The limiting value of $\tilde{\gamma}(L)$ is known in the theory of products of random matrices\cite{random matrices} as the Lyapunov exponent (LE). The inverse quantity, the localization length, $l_{loc}=\gamma^{-1}$, determines the main length scale in the localization regime.\cite{LGP} SPS in this context means that the variance of $\tilde{\gamma}$ is not an independent parameter, but it is determined by $\gamma$ itself implying a simple relationship between the two\cite{Anderson} 
\begin{equation}
\sigma ^2 = \gamma /L .
\label{SPS}
\end{equation}
This expression was obtained for the one-dimensional model in Ref. \onlinecite{Anderson} assuming complete randomization of the phases of complex transmission and reflection coefficients over a microscopic length scale $l_{ph}\ll l_{loc}$ (phase randomization hypothesis). Later the phase randomization hypothesis was used by many different authors to rederive Eq. (\ref{SPS}) (see, for instance, Refs. \onlinecite{LGP,Stone}), and the inequality $l_{ph}\ll l_{loc}$ came to be regarded as the criterion for SPS.  However, there were earlier signs that the phase randomization hypothesis is neither a necessary nor a sufficient condition for SPS to occur. For instance, numerical simulations of Ref. \onlinecite{Stone} and analytical calculations of Ref.\onlinecite{Dmitriev} showed that in the center of a conductivity band of the 1-$D$ Anderson model, SPS holds even though the phase is not randomized, provided that the disorder is weak. Later, numerical simulations of a random, periodic-on-average model\cite{Dima} demonstrated  a strong violation of  SPS in the band-gaps of the spectrum of the underlined system without disorder, which existed even for weak disorder and, actually, diminished, when disorder increased. This was contrary to the behavior found for states from the original conducting band, for which minor deviations of the variance from Eq. (\ref{SPS}) occured when disorder becomes strong enough.\cite{Stone}

The final realization of the fact that the phase randomization hypothesis has nothing to do with SPS came in Ref. \onlinecite{Altshuler}. In that paper, the variance of LE was calculated exactly for the Lloyd model and the SPS equation (\ref{SPS}) was derived without {\it ad hoc} assumptions. It was found that the emergence of SPS is governed by a new length scale $l_s$, related to the integral density of states with the new criterion for SPS being $\kappa =l_{loc}/l_{s}\gg 1$. On the basis of the exact solutions it was conjectured\cite{Altshuler} that in the region of the spectrum close to its original boundary, the parameter $l_s$ can be defined in a generic case as 
\begin{equation}
l_s=a/N(E),\label{ls}
\end{equation} 
where  $a$ is the lattice constant and $N(E)$ is the number of states between the lowest genuine  boundary of the spectrum of the disordered system and $E$, normalized by the total number of states in the band (such that $0<N(E)<1$). Numerical studies undertaken in Ref. \onlinecite{Altshuler} evidenced that this assumption was valid for the Anderson model and the  periodic-on-average model with rectangular distribution of random parameters. An additional implication of the results of Ref. \onlinecite{Altshuler} is that the random matrix theory approach, which also reproduces Eq. (\ref{SPS}),\cite{Mello,Beenakker1} does not apply to spectral regions with depleted differential density of states.

The objective of this paper is to show that two different scaling regimes governed by the parameter $l_s$ found in Ref.\onlinecite{Altshuler} exist also in disordered systems with absorption. Having in mind applications to light propagation in random photonic band gap materials, we consider how the inclusion of small absorption affects scaling properties of transmission. We show that the conjectured definition of $l_s$ in terms of the integral density of states can also be applied to disordered systems with absorption. 

Within the phase randomization hypothesis approach, statistics of the transmission in lossy one-dimensional dielectrics was considered analytically in a number of papers.\cite{Freilikher,Beenakker2} Following Ref. \onlinecite{Freilikher} the relation between the variance and the localization length in the presence of absorption can be presented in the form 
\begin{equation}
\tau=\tau_0 (\beta)=1+2\beta e^{2\beta} Ei(-2\beta),
\label{SPS_abs}
\end{equation}
where $\tau =\sigma^2 L/\gamma$, $\beta =\l_{loc} /l_{a}$, and $l_a$ is the absorption length in the absence of randomness,
$$
Ei(x)=\int_{-\infty}^{x} dt \; exp(t)/t
$$
is the exponential integral. In the absence of absorption, Eq. (\ref{SPS_abs}) reduces to the regular SPS form  $\tau =1$. In two limiting cases Eq. (\ref{SPS_abs}) gives the asymptotes $\tau = 1-2\beta \ln(1/\beta), \ \ \beta \ll 1$ and  $\tau = 1/2\beta, \ \ \beta \gg 1$. 

Presuming that the length scale $l_s$ retains its meaning in the case under consideration, we expect that the parameter $\tau $ deviates from the phase randomization hypothesis prediction $\tau_0 (\beta )$ [Eq. (\ref{SPS_abs})] in the vicinity of the boundaries of the spectrum in accordance with the same criterion $l_{loc}\gg l_s$ as in Ref. \onlinecite{Altshuler}. Using numerical simulations of a periodic-on-average one-dimensional lossy system, we show that, indeed, the parameter $\kappa=l_{loc}/l_s$ sets a valid criterion for validity of Eq. (\ref{SPS_abs}). As an additional benefit, we demonstrate that within the range of its validity, Eq. (\ref{SPS_abs}) represents a universal, model independent relation between the variance and the mean value of LE. Though the statistics of transmission is determined now by two parameters, one can still regard Eq. (\ref{SPS_abs}) as a generalization of SPS because the absorption length is independent of disorder. The deviations from the phase-randomization based results of Ref.\onlinecite{Freilikher} studied in our paper must be clearly distinguished from results of Ref.\onlinecite{Gupta}. In the latter paper brake down of the phase randomization was obtained  in the case of very strong disorder and strong absorption for states at the center of the original band. The results of our paper indicate that (i) violation of the generalized single parameter scaling occurs at weak disorders for states close to the band edge of the original spectrum, (ii) this violation is not related to the phase randomization but is controlled by the parameter $l_s$.  

\section{Model and the method of calculations}

We consider a classical transverse electromagnetic wave propagating normally through a stack of alternating dielectric slabs with dielectric constants $\epsilon _1$ and $\epsilon _2$. The widths of the stacks of the first kind is distributed uniformly in the interval $(d_1 -\delta ,d_1 +\delta)$ while the width of the others is being kept constant $d_2$. The propagation of the waves in the superlattice consists of free propagation in the slabs and scattering at the interfaces, where the boundary conditions should be satisfied. It can be described using the transfer matrix formalism for the vector $v_n =(E_n, E^{'}_n /k)$, where $E_n, E^{'}_n$ are the electric field and its derivative at $n$-th interface and $k=\omega /c$. The presence of absorption can be accounted for by adding a constant complex part to the dielectric functions: $\epsilon _1=\epsilon _1^{(0)}(1+i\alpha )$ and $\epsilon _2=\epsilon _2^{(0)}(1+i\alpha )$, where $\alpha $ is a damping coefficient. Vectors on neighboring interfaces are connected via the transfer matrix
\begin{equation}
T _{n}=
\left( 
\begin{array}{cc}
\cos k_n d_n  & 
\frac{1}{k_n} \sin k_n d_n \\ 
- k_n \sin k_n d_n & 
\cos k_n d_n  
\end{array}
\right) ,  
\label{transfer matrix}
\end{equation}
where $k_n=k\sqrt{\epsilon _n}$. The transfer matrix of the entire system is ${\bf T} (\alpha ,L)=\Pi_{n=1}^{2N} T_n(\alpha )$, where $L=N(d_1+d_2)$. LE is defined through the transmission coefficient for the superlattice:
\begin{equation}
\gamma (L,\alpha ) = -\frac{1}{4 N} \left< \ln (t t^{*}) \right>  = -\frac{1}{4 N} \left< \ln \left| \frac {2 \det {\bf T}(\alpha ,L) }{({\bf T} _{11}(\alpha ,L)+{\bf T} _{22}(\alpha ,L))-i({\bf T} _{12}(\alpha ,L)-{\bf T} _{21}(\alpha ,L)) } \right| ^2 \right> ,
\label{lyap1}
\end{equation}
here $\left< ... \right>$ denotes the average over an ensemble of configurations.  We find that using this definition in numerical simulations has one significant shortcoming. In long systems  , the transmission coefficient $t t^{*}$ falls bellow the computer roundoff accuracy. The usual remedy for this problem in the absence of absorption is to use an alternative definition of $\gamma$ 
\begin{equation}
\gamma(L,\alpha ) = \frac{1}{4 N} \left< \ln \frac {|| {\bf T}(L) v_0 ||^2}{|| v_0 ||^2}\right>,
\label{Taltern}
\end{equation}
which allows one to consider very long systems.\cite{random matrices} In the presence of absorption, this definition must be generalized because the simple substitution of the transfer-matrix, Eq. (\ref{transfer matrix}), in this equation would lead to a wrong result. The problem is that an exponentially growing eigenvalue of the transfer matrix appears in the denominator of Eq. (\ref{lyap1}), while in Eq. (\ref{Taltern}) it is in the numerator. Therefore, the contribution from absorption enters the final answer for the LE with different relative signs in these two equation. We argue that  Eq. (\ref{Taltern}) must be modified in order to agree with the original definition of ${\gamma}$ as follows:
\begin{equation}
\gamma (L,\alpha ) = \frac{1}{4 N} \left< \ln \frac {|| {\bf T}(-\alpha ,L) v_0 ||^2}{|| v_0 ||^2}\right>  ,
\label{lyap2}
\end{equation}
where $v_0$ is a vector of general position. The proof of equivalency of Eq. (\ref{lyap1}) and Eq. (\ref{lyap2}) goes as follows. First, we diagonalize the transfer matrix ${\bf T}(\alpha ,L)$:
\begin{equation}
{\bf T}(\alpha ,L)=U^{\dagger} \ {\bf T}^{(D)}(\alpha ,L)\ U =U^{\dagger} 
\left( 
\begin{array}{cc}
e^{\nu _1(\alpha )2N}& 0 \\ 
0 & e^{\nu _2(\alpha )2N}
\end{array}
\right) U,  
\label{diag}
\end{equation}
here $U$ and $U^{\dagger}$ are some unitary matrices. For argument sake, we assume that 
$|\nu _1(\alpha )|\geq|\nu _2(\alpha )| $. Next, we notice the following relation between eigenvalues of the transfer matrix
\begin{equation}
\nu _1(\alpha )=-\nu _2(-\alpha ),
\label{recip}
\end{equation}
which follows simply from reciprocity of our system. Indeed, propagation of the waves in opposite direction should be described by the matrix ${\bf T}^{'}(\alpha ,L)=\left[ {\bf T}(-\alpha ,L) \right] ^{-1}$, that leads to Eq (\ref {recip}). Now, substituting ${\bf T}^{(D)}(\alpha ,L)$ into Eqs. (\ref{lyap1}) and (\ref{lyap2}) (corrections due to matrices $U,U^{\dagger} $ are negligible in the limit $N\rightarrow \infty$) one obtains $\gamma (L,\alpha )=-\nu _2(\alpha )$ and $\gamma (L,\alpha )=\nu _1(\alpha )$ respectively. Taking into account the reciprocity relation Eq. (\ref{recip}) we establishes the equivalency of Eqs. (\ref{lyap1}) and (\ref{lyap2}). 

\section{Results}

\begin{figure}
\centering
\vspace{-0.2in}
\epsfxsize=4in \epsfbox{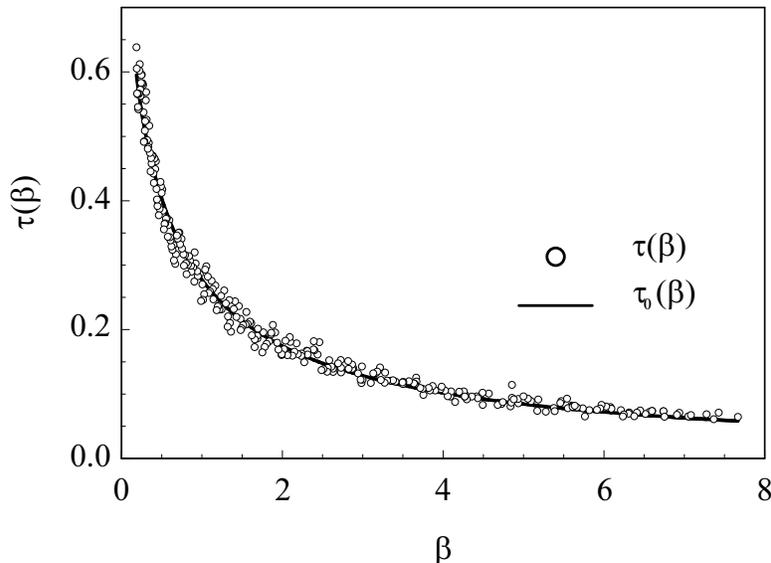}
\caption{Open circles depict the parameter $\tau (\beta )$ computed for the periodic-on-average system of 100,000 layers far from the band edge. $0.86<k<1.4$, the disorder parameter is $\delta =0.45$, the  damping $\alpha =0.00125$. The solid line shows $\tau _0(\beta)$ given by Eq. (\ref{SPS_abs})}
\end{figure}

In numerical simulations we used the following set of parameters $\epsilon_1=1.2$, $\epsilon_2=1$, $\left< d_1 \right> =d_2 =1$. The disorder parameter, $\delta $, and the absorption rate, $\alpha $, were variable parameters. To calculate the moments of $\gamma $, we averaged the characteristics of systems as long as $100,000$ layers over $5000$ realizations. The size of the stack was chosen to be at least five times the localization length or the absorption length. In the ordered system the first forbidden gap lies between $k=1.456$ and $k=1.543$. Since the localization length depends on the frequency of the wave it is possible to study the function $\tau (\beta (\omega ) )$ by changing the frequency. First, we compare our numerical results with the analytical formula of Ref. \onlinecite{Freilikher} in the region of frequencies well inside the first allowed band, where we expect these results to coincide.   Fig. 1. shows excellent agreement between the computed $\tau (\beta )$ for $0.86<k<1.4$  and Eq. (\ref{SPS_abs}).
\begin{figure}
\centering
\vspace{-0.1in}
\epsfxsize=4in \epsfbox{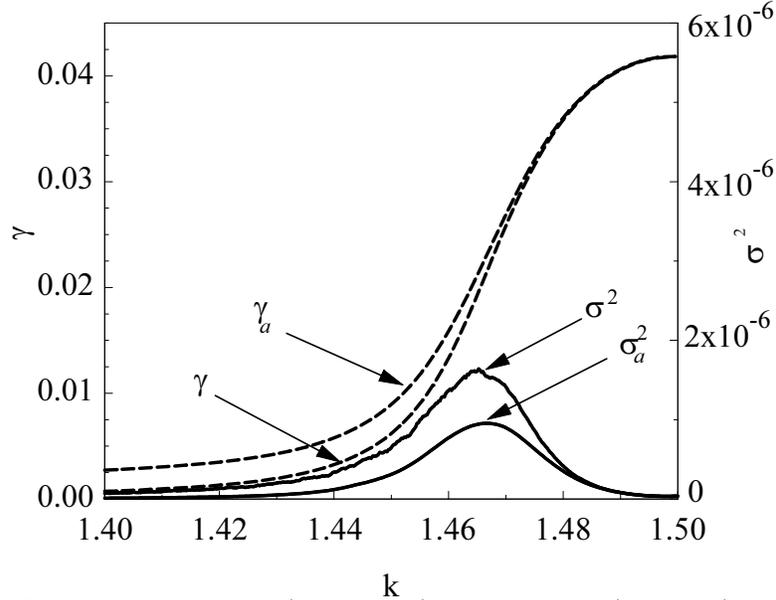}
\caption{The dependence of the lyapunov exponent (dashed lines) and its variance (solid lines) averaged over 2500 realizations with and without absorption as function of the frequency on the band edge. The length of the system 10,000, disorder parameter $\delta =0.25$, damping $\gamma =0.0025$.}
\end{figure}

The localization length decreases rapidly when the frequency approaches the band edge, while LE, consequently, increases. As follows from Eq. (\ref{SPS}), in the absence of absorption, $\sigma ^2$ should follow the LE. Fig. 2 depicts the dependence of LE and its variance with and without absorption. As one can see, at the band edge, $k=1.456$, SPS breaks down. Indeed, while LE grows with the increasing frequency, the variance drops in both cases, with and without absorption. 
\begin{figure}
\centering
\vspace{-0.0in}
\epsfxsize=4in \epsfbox{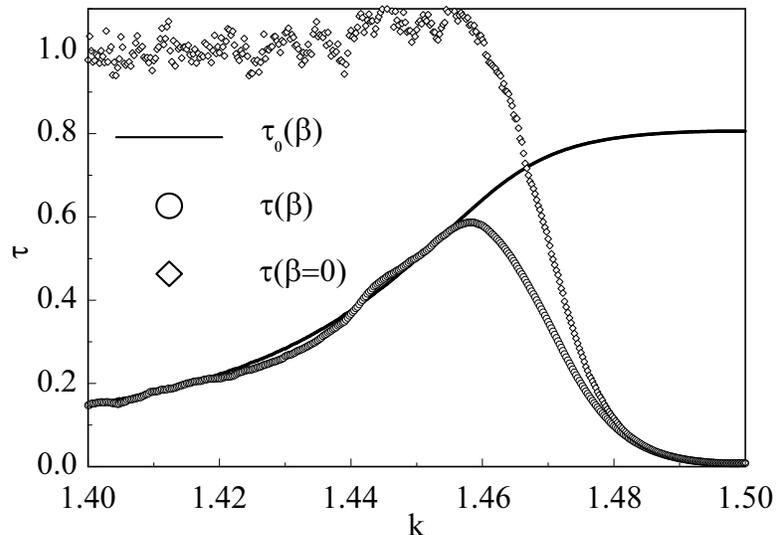}
\caption{$\tau (0)$ (crosses), $\tau (\beta)$ (open circles) and $\tau _0(\beta)$ (solid line) on the band edge plotted as function of frequency for the same set of parameters as on Fig. 2}
\end{figure}

In order to compare numerical and analytical results in the presence of absorption, we calculated the parameter $\tau $ as a function of frequency $k$, using the data presented in Fig. 2. The results are presented in Fig. 3, from which one can see that the computed $\tau (\beta)$ deviates from  $\tau_0 (\beta)$ that represents Eq. (\ref{SPS_abs}), and that this deviation occurs at  the same frequency at which $\tau (0)$  deviates from unity. These graphs convincingly demonstrate, that even in the presence of absorption, the spectrum of the system is separated in groups with different scaling properties, and that the boundary between the groups coincide with the  boundary of the original spectrum.  

The next question we need to address is whether the transition between the scaling regimes is governed by the same parameter $l_s$ that was introduced in Ref. \onlinecite{Altshuler}. According to that paper, the new parameter $l_s$ becomes greater than the localization length at the band edge, which results in a deviation from SPS. Using definition of $l_s$ suggested in Ref. \onlinecite{Altshuler} regarding the integrated density of states, Eq. (\ref{ls}), we numerically calculated this parameter for the system studied in the present paper. The density of states was calculated using the phase formalism (see, for instance, Ref. \onlinecite{LGP}) for the system without absorption. Fig. 4 shows all relevant length parameters: localization length, absorption length and $l_s$.
\begin{figure}
\centering
\vspace{0.2in}
\epsfxsize=4.5in \epsfbox{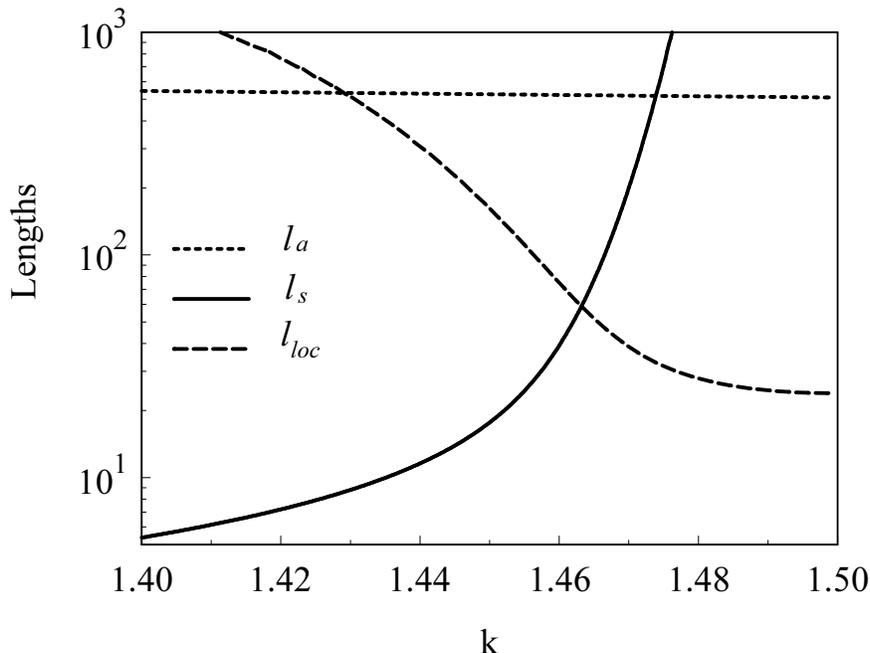}
\caption{Length scales for the system with parameters used on Fig. 2,3. Solid line depicts $l_s$, dashed line is the localization length $l_{loc}$, and dotted line is the absorption length $l_a$}
\end{figure}

At the band edge, $l_s$ grows rapidly because very few new states appear within the former bandgap, and $N(E)$ must already be close to unity at the band edge. It  reaches unity at a new fluctuation boundary of the spectrum near the center of the gap. We assume that the disorder is not very strong such that the fluctuation boundaries inside former bandgaps do exist. If the disorder is strong enough, or if its statistical properties are such that the entire bandgap is filled with fluctuation states a definition of $l_s$ is still possible, but the situation becomes more complicated, and we do not consider it here.  At certain points $l_s$ grows larger than the localization length $l_{loc}$, and one can find comparing Figs. (3) and (4), that $\tau (\beta)$ starts deviating from $\tau_0 (\beta)$ at the same frequency. In order to make this more transparent, it is convienient to plot various $\tau$'s versus $\kappa =l_{loc}/l_s$.

Without absorption, we see the crossover to SPS in its pure form at $\kappa \simeq 1$ reported in Refs. \onlinecite{Dima,Altshuler}. When absorption is present, the crossover still occurs at  $\kappa \simeq 1$, but now to the modified SPS behavior $\tau _0(\beta)$. We stress that the crossover occurs while $a \ll l_{loc}$, $a \ll l_{a}$, $l_{loc} \ll L$, and $l_a \ll L$, so that the system remains in the meaningful scaling regime. This demonstrates that the condition $1\ll \kappa $ establishes the  criterion for modified SPS even in the presence of the absorption. 

To check that our results are not model specific we also studied the Anderson model, with absorption introduced as a nonrandom imaginary part of the on-site random energy. In the Anderson model without disorder and absorption, there is a single allowed band form $E=-2$ to $E=2$. 
\begin{figure}
\centering
\vspace{-0.1in}
\epsfxsize=4.5in \epsfbox{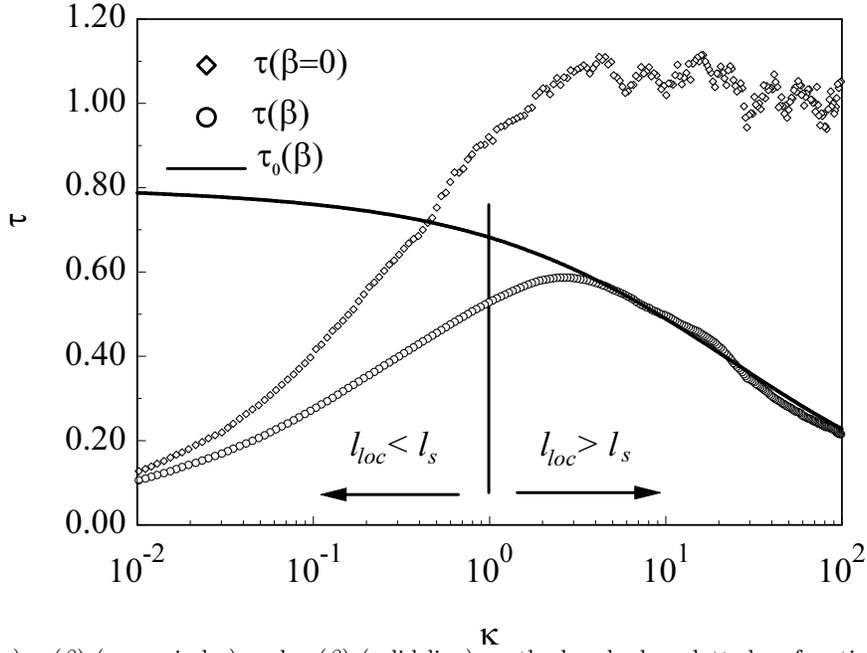}
\caption{$\tau (0)$ (crosses), $\tau (\beta)$ (open circles) and $\tau _0(\beta)$ (solid line) on the band edge plotted as function of $\kappa =l_{loc}/l_s$ for the same set of parameters as on Fig. 2,3}
\end{figure}

Figure 6 shows that $\tau (\beta)$ deviates from $\tau_0(\beta)$  at the edges of the allowed band as would be expected on the basis  of the results presented for the periodic-on-average system. Inside the conductivity band, numerical and analytical results show excellent agreement. This is interesting in itself, since Eq. (\ref{SPS_abs}) was derived for a continuous model, \cite{Freilikher} and we see that in the SPS regime it holds  also for the tight-binding discrete model. 

\begin{figure}
\centering
\vspace{-0.1in}
\epsfxsize=4.5in \epsfbox{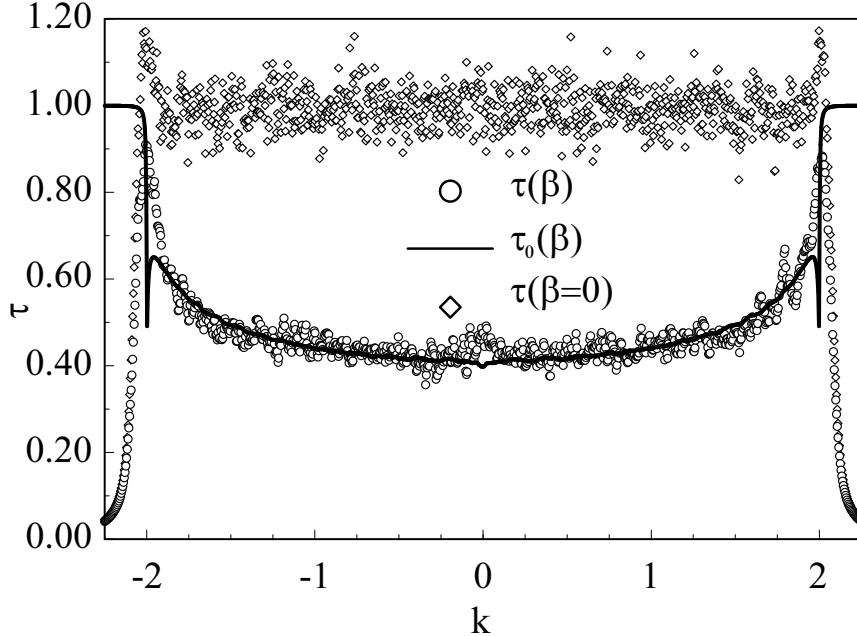}
\caption{$\tau (0)$ (crosses), $\tau (\beta)$ (open circles) and $\tau _0(\beta)$ (solid line)  plotted as function of frequency averaged over 1000 realizations. The length of the system 10,000 cells, disorder parameter $\delta =0.25$, damping $\gamma =0.0025$.}
\end{figure}

In conclusion, we computed the Lyapunov exponent and its variance for a periodic-on-average layered system and for the one-dimensional Anderson model. We studied the deviation from the absorption-modified SPS expression for the variance.\cite{Freilikher} We showed that the new length scale, introduced in Ref. \onlinecite{Altshuler} in order to explain violation of SPS in the systems without absorption, retains its significance when absorption is present. We also showed that the same criterion $ \kappa =l_{loc}/l_s \gg 1$ derived for lossless systems in Ref. \onlinecite{Altshuler} controls scaling behavior in  lossy systems.

We are indebted to A. Genack for reading and commenting on the manuscript. This work was partially supported by NATO Linkage Grant N974573, CUNY Collaborative Grant, and PSC-CUNY Research Award.

%\end{multicols}

\end{document}